\documentclass[10pt,letterpaper]{article}
\usepackage[top=0.85in,left=2.75in,footskip=0.75in]{geometry}

\usepackage{amsmath,amssymb}
\usepackage{mathtools,xparse}
\usepackage[ruled,vlined]{algorithm2e}
\usepackage{algorithmic}
\newtheorem{model}{Model}

\usepackage{changepage}
\usepackage{setspace}
\usepackage{subcaption}

\usepackage[utf8x]{inputenc}

\usepackage{textcomp,marvosym}

\usepackage{cite}

\usepackage{nameref,hyperref}

\usepackage[right]{lineno}

\usepackage{microtype}
\DisableLigatures[f]{encoding = *, family = * }

\usepackage[table]{xcolor}

\usepackage{array}

\newcolumntype{+}{!{\vrule width 2pt}}

\newlength\savedwidth

\raggedright
\usepackage[aboveskip=1pt,labelfont=bf,labelsep=period,justification=raggedright,singlelinecheck=off]{caption}

\makeatletter
\renewcommand{\@biblabel}[1]{\quad#1.}
\makeatother

\usepackage{lastpage,fancyhdr,graphicx}
\usepackage{epstopdf}
\fancyheadoffset[L]{2.25in}
\fancyfootoffset[L]{2.25in}
\DeclareMathOperator{\diag}{diag}

\DeclarePairedDelimiter{\norm}{\lVert}{\rVert}

\begin{document}
\vspace*{0.2in}

\begin{flushleft}
{\Large
\textbf\newline{Multiple two-sample testing under arbitrary covariance dependency with an application in imaging mass spectrometry} 
}
\newline
\\
Vladimir Vutov\textsuperscript{1},
Thorsten Dickhaus\textsuperscript{1*}
\\
\bigskip
\textbf{1} Institute for Statistics, University of Bremen, Bremen, Germany
\\
\bigskip

%
%





* dickhaus@uni-bremen.de

\end{flushleft}
\section*{Abstract}
Large-scale hypothesis testing has become a ubiquitous problem in high-dimensional statistical inference, with broad applications in various scienfitic disciplines. One relevant application is constituted by imaging mass spectrometry (IMS) association studies, where a large number of tests are performed simultaneously in order to identify molecular masses that are associated with a particular phenotype, e.\ g., a cancer subtype. Mass spectra obtained from Matrix-assisted laser desorption/ionization (MALDI) experiments are dependent, when considered as statistical quantities. False discovery proportion (FDP) control under arbitrary dependency structure among test statistics is an active topic in modern multiple testing research. In this context, we are concerned with the evaluation of associations between the binary outcome variable (describing the phenotype) and multiple predictors derived from MALDI measurements. We propose an inference procedure in which the correlation matrix of the test statistics is utilized.  
The approach is based on multiple marginal models (MMM). Specifically, we fit a marginal logistic regression model for each predictor individually. Asymptotic joint normality of the stacked vector of the marginal regression coefficients is established under standard regularity assumptions, and their (limiting) correlation matrix is estimated.  
The proposed method extracts common factors from the resulting empirical correlation matrix.  Finally, we estimate the realized FDP of a thresholding procedure for the marginal $p$-values.  We demonstrate a practical application of the proposed workflow to MALDI IMS data in an oncological context.


\section*{Introduction}
Imaging mass spectrometry (IMS) is a technique that acquires spatially resolved mass spectral information of small to large molecules. Provided a thin tissue section, mass spectra are collected in a spatially orientated pattern within the tissue. This produces an image, where each discrete spot represents a mass spectrum. Mass spectra associate molecular masses to their relative molecular abundances. Hence, this provides insights into the chemical decomposition of a unique and specific region in the tissue.
A promising technology that has evolved over the recent years is Matrix-assisted laser desorption/ionization (MALDI) imaging mass spectrometry, also known as MALDI imaging. This technology allows for analysing a wide range of analytes (e.g. proteins, peptides, lipids, etc.) from many types of biological samples. MALDI imaging is a versatile tool and has the advantage of combining spatial and molecular information from biological samples. This makes the technology interesting for biomedical and cancer research (for more pathological applications, see e.g. \cite{aichler}, \cite{krieg}). The latter is possible by virtue of its applicability to analysing formalin-fixed paraffin-embedded (FFPE) tissue samples. 
One of the key benefits of utilising MALDI IMS on fixed samples is that multiple FFPE core biopsies can be arranged in a single tissue microarray (TMA) block (see, \cite{pote}, \cite{Boskamp}).
 Hence, within an individual run of the mass spectrometer a cohort of possibly cancerous tissues can be examined simultaneously in order to extract biochemical information spatially. 
Respectively, such biochemical information can be used for the determination of the cancer subtypes or the identification of the origin of the primary tumour in patients with metastatic disease.
 Accurate typing of a tumour is an essential requirement for the successful treatment of patients.  
As pointed out in \cite{Boskamp}, modern MALDI-IMS instruments manage to acquire molecular information with a small signal-to-noise ratio at short time measurements.
 
 A challenging task for advanced bioinformatics tools, as acknowledged in \cite{alex}, is stable feature extraction or, in other words, extracting biologically meaningful evidence out of a huge amount of spectra.
  A common practice for identifying meaningful features relies on the idea of discovering considerable signal peaks, which is also known as  peak detection. These peaks, respectively, are anticipated to be distinctive for cancer identification.
  Statistically, we model each spectrum individually as it is measured from small tissue core regions with slight fluctuating structure within a single core.  
 
Large-scale multiple testing is a widely used methodology in the analysis of high-dimensional data and has a variety of applications in scientific fields like, e.\ g., genomics, proteomics, brain-computer interfacing, etc. (for more life science applications, cf. Chapters 9-12 in \cite{dickhausbook}). Starting with the highly influential work \cite{BH}, control of the expected proportion of false positive findings, called false discovery rate (FDR), has become a standard type I error criterion in large-scale multiple testing. 
 Another well-known technique to control the FDR has been proposed by \cite{storey2002}, and is often referred to as Storey's procedure. Its main idea is to fix a rejection threshold value $t$ for the marginal $p$-values, then to estimate the FDR of the resulting thresholding procedure, and finally to choose ${t}$ such, that the estimated FDR is lower than or equal to the pre-defined FDR level $\alpha$. Early FDR research has mainly established FDR control of the aforementioned procedures in the case of independent test statistics. However, high-dimensional studies seldom involve the analysis of independent variables. In contrast, most studies involve many related variables simultaneously (cf., among many others, \cite{madam}, \cite{friquet2009}). Similarly, MALDI-IMS data consist of a couple of thousands of variables, and many of them are related.  Explicitly taking into account these dependencies can increase the power of the multiple test, cf. \cite{handbook-chapter} for an overview of so-called multivariate multiple tests. 
 
There are multivariate multiple tests which are based on block structures in the data. For instance, in \cite{madam} it has been proposed to control the family-wise error rate (FWER) in blocks of adjacent genetic markers; see also Section 5 in \cite{MVCHS}. Likewise, in \cite{block} an extensive study to compare different controlling methods based on the assumption of block-correlation positively dependent tests has been reported. However, there is no evidence that MALDI-IMS data can be grouped straightforwardly into adjacent blocks. 
 Other methods utilize a multi-factor model in order to describe the  dependencies among the test statistics, meaning that the latter dependency structure may be explained by latent factors.

In addition to modelling the dependencies, a further task is to integrate the correlation effects in the decision process; see, for example, \cite{efron2007, efron2010, leek}. 
In \cite{pfa}, a general setting for approximating the false discovery proportion (FDP) has been introduced. The authors assumed that the test statistics are (approximately) following a multivariate normal distribution with an arbitrary and known covariance matrix. The idea of their approach is to carry out a spectral decomposition of the covariance matrix of the test statistics, and then to subtract the principal factors that cause the strong dependency across the z-values before evaluating  the FDP. This method is called principal factor approximation (PFA). 
In \cite{fan_poet},  a fully data-driven process to estimate the FDP has been established, where the authors adopted a POET estimator (see \cite{poet}) to estimate an unknown covariance matrix, and subsequently to compute the realized FDP.
Recently, in \cite{farmtest} the problem when the assumption of normality is violated has been addressed, for instance in the context of multiple testing under arbitrary dependency and heavy-tailed data. The method utilizes a robust covariance estimator and constructs factor-adjusted test statistics.

In the present work, we explore the problem of two-sample multiple hypotheses testing under arbitrary correlation dependency under the scope of multiple marginal logistic regression models by making use of PFA. Furthermore, we apply our proposed method to MALDI Imaging data.  

\section*{Materials and methods}
In MALDI imaging related studies, data are commonly stored in an $n \times p$ matrix $X = (x_{ij})_{\substack{1 \leq i \leq n \\ 1 \leq j \leq p}}$, where spectra are stored as rows and columns correspond to mass-to-charge (m/z) values (in the context of MALDI interpreted
as the molecular mass, cf.  \cite{alex}). Usually, both $n$ and $p$ are in the thousands. 
We address the biological question of the association between m/z values and a cancerous status by testing multiple hypotheses. More specifically, we test for each $j$ the null hypothesis $H_{j}$ which states there is no association between a particular m/z value $X_{j}$ and the cancer subtype.
\subsection*{Marginal Modelling} 
The first step of the proposed framework is to model the marginal associations, for each $j$ separately. Therefore, let $j$ be arbitrary, but fixed throughout this section. Let $Y$ denote the (random) binary outcome and let $X_{j}$ denote the random variable describing the $j$-th m/z value. Thus, the tuple $(X_{j}, Y)$ takes its values in $\mathbb{R} \times \{0, 1\}$. We are interested in the conditional distribution $\mathbb{P}(Y| X_j)$. To this end, we assume a (marginal) binary regression model with the canonical (logit) link function. This model has two parameters, namely, the intercept $\alpha_j$ and the regression coefficient $\beta_j$. We denote the observational units for the $j$-th marginal regression problem by $(X_{j}^{(i)}, Y^{(i)})_{1 \leq i \leq n}$, and we assume that they are independent copies of $(X_{j}, Y)$. 
Letting, for a given $i \in \{1, \ldots, n\}$, $\pi_{j}^{(i)} = \mathbb{P}(Y^{(i)} = 1 | X_j^{(i)})$, the model equation for the $j$-th binary logistic regression models is given by
\begin{equation}
\label{link}
g(\pi_{j}^{(i)}) := \log \left( \frac{\pi_{j}^{(i)}}{1 - \pi_{j}^{(i)}} \right) = \alpha_{j} + X_{j}^{(i)}\beta_{j},
\end{equation}
where $g = \text{logit}$ is the canonical link function mentioned before.
The unknown parameters $(\alpha_{j}, \beta_{j})$ are estimated by the principle of the maximum (log-) likelihood. The log-likelihood function pertaining to the model in \eqref{link} is given by
\begin{align} 
\label{mle}
{l}(\alpha_{j}, \beta_{j}) &= \sum_{i = 1 }^{n} Y^{(i)}\left[\log \pi^{(i)}_{j} - \log (1 - \pi^{(i)}_{j})\right] + \log(1 - \pi^{(i)}).
\end{align}
By substituting 
\[
\pi_{j}^{(i)} = \frac{\exp(\alpha_{j} + X_{j}^{(i)}\beta_{j})}{1 + \exp(\alpha_{j} + X_{j}^{(i)}\beta_{j})} \text{~~as well as~~} 1 - \pi^{(i)}_{j} = \frac{1}{1 + \exp(\alpha_{j} + X_{j}^{(i)}\beta_{j})}
\] 
in \eqref{mle}, we obtain that
\begin{equation}\label{log_model}
l(\hat{\alpha}_{j}, \hat{\beta}_{j}) = \max_{(\alpha_{j}, \beta_{j})}  \sum_{ i = 1}^{n} Y^{(i)} \left(\alpha_{j} +X_{j}^{(i)}\beta_{j}\right) - \log\left(1 + \exp(\alpha_{j} + X_{j}^{(i)}\beta_{j})\right),
\end{equation}
where the estimation is performed conditionally to the actually observed values $X_{j}^{(i)} = x_{j}^{(i)}$ for $1 \leq i \leq n$.

In this study, we are concerned with simultaneous testing of the pairs of hypotheses
\begin{equation}
\label{testing}
H_{0j}: \beta_{j} = 0 \text{~~versus~~} H_{1j}: \beta_{j} \neq 0, \quad j = 1, \cdots, p.
\end{equation}
Biologically speaking, we aim at discovering the most distinctive m/z values for a cancer association.
\subsection*{Multiple Marginal Models}
The second step of the proposed procedure is to combine all $p$ marginal models and to approximate the joint null distribution of all estimators.
To this end, we follow the framework described in \cite{mmmee} for jointly estimating multiple marginal association parameters, and apply this framework to the marginal models described in the previous section. Notice that we assume that regression
coefficients are unique to one model $j$ and not shared between any two models $j_1 \neq j_2$. Furthermore, the intercepts $(\alpha_j)_{1 \leq j \leq p}$ are nuisance parameters in the sense that the hypotheses in \eqref{testing} only refer to the $\beta_j$'s. 

The main goal of this section is to establish a central limit theorem for the vector $\hat{\beta} = ( \hat{\beta}_{1}, \cdots, \hat{\beta}_{p})^\top$, which is achieved by stacking the score contributions of the $\hat{\beta}_{j}$'s across all $p$ marginal models. Following \cite{mmmee}, we consider the asymptotic ($n \to \infty$) expansion
\begin{equation}\label{asym-univ}
(\hat{\beta}_{j} - \beta_{j}) \sqrt{n} = \frac{1}{\sqrt{n}} \sum_{i = 1}^{n} \Psi_{ij} + o_{\mathbb{P}}(1),
\end{equation}
where $\Psi_{ij} = F(\beta_{j})^{-1}\widetilde{\Psi}_{ij}$,
$F(\beta_{j})^{-1}$ is the appropriate row of the inverse Fisher information matrix that corresponds to $\beta_{j}$, $\widetilde{\Psi}_{ij}$ is the score function (the first derivative of the log-likelihood function) for the $i$-th observational unit, and $o_{\mathbb{P}}(1)$ indicates a sequence of random variables converging to zero in probability.

Now, we define the vectors
 $\beta := ({\beta}_{1}, \cdots, \beta_{p})^\top$, $\hat{\beta} := (\hat{\beta}_{1}, \cdots, \hat{\beta}_{p})^\top$, and $\Psi_{i} := (\Psi_{i1}, \cdots, \Psi_{ip})^\top$, and consider the asymptotic expansion  
\begin{equation}
\label{asym}
(\hat{\beta} - \beta) \sqrt{n} = \frac{1}{\sqrt{n}} \sum_{i = 1}^{n} \Psi_{i} + o_{\mathbb{P}}(1),
\end{equation}
which follows from \eqref{asym-univ} under standard regularity assumptions like, for instance, finiteness of the Fisher information and non-vanishing (limiting) proportion of data points corresponding to $Y=1$ and $Y=0$, respectively. The left-hand side of \eqref{asym} converges in distribution, by the multivariate central limit theorem, to a $p$-variate normal distribution, i.\ e., 
\begin{equation}
(\hat{\beta} - \beta) \sqrt{n} \xrightarrow{d} N_p(0, \Sigma).
\end{equation}
The limiting variance-covariance matrix $\Sigma$ can be estimated in a consistent manner, namely by
\begin{equation}
\label{mmmvariance}
\widehat{\Sigma} = \frac{1}{n} \sum_{i = 1}^{n} \hat{\Psi}_{i}^{\top}\hat{\Psi}_{i},
\end{equation}
whence  $\hat{\Psi}_{i}$ are yielded by plugging the parameter estimates from all marginal models into $\Psi_{i}$.

In our study, we are genuinely interested in the effect of $\beta_{j}$ for $j \in \{1, \ldots, p\}$. However, the intercepts $(\alpha_j)_{1 \leq j \leq p}$ contribute to the estimation and standardisation of the $\beta_{j}$'s. Specifically, for the logit model described in the previous section, $\hat{\Psi}_{ij}$ is given by the second coordinate of the bivariate vector
\begin{equation}
\left\{\hat{\pi}_{j}^{(i)}(1-\hat{\pi}_{j}^{(i)}) (1, X^{(i)}_{j})^\top (1, X^{(i)}_{j})\right\}^{-1} (1, X^{(i)}_{j})^\top (Y^{(i)} - \hat{\pi}_{j}^{(i)}),
\end{equation}
where  $\hat{\pi}_{j}^{(i)} = \frac{\exp(\hat{\alpha}_{j} + X_{j}^{(i)}\hat{\beta}_{j}))}{1 + \exp(\hat{\alpha}_{j} + X_{j}^{(i)}\hat{\beta}_{j})}$ and $\hat{\alpha}_{j}$,	 $\hat{\beta}_{j}$ are as in \eqref{log_model}.
 
Next, we denote  by $Z_1, \ldots, Z_{p}$ the Studentized versions of $\hat{\beta}_{1}, \ldots \hat{\beta}_{p}$, meaning that
\begin{equation}
\label{z_values}
Z_j = \frac{\hat{\beta}_{j}}{\sqrt{\widehat{\text{Var}}({\hat{\beta}_j})}}, \quad j = 1, \cdots, p,
\end{equation}
where $\sqrt{\widehat{\text{Var}}(\hat{\beta}_j)}$ is the square root of the $j$-th diagonal element of $\hat{\Sigma}$, divided by $\sqrt{n}$. 
Then, we have that
\begin{equation}
\label{zstat}
(Z_{1}, Z_{2}, \ldots, Z_{p})^{\top} \underset{\text{approx.}}{\sim} N_p((\mu_{1}, \mu_{2}, \ldots, \mu_{p})^{\top}, \hat{\Sigma}^*),
\end{equation}
where $\mu_{j} = \beta_{j} / \sqrt{\widehat{\text{Var}}(\hat{\beta}_j)}$ for $1\leq j \leq p$, $\hat{\Sigma}^* = \diag[\hat{\Sigma}]^{-1/2} \hat{\Sigma}   \diag[\hat{\Sigma}]^{-1/2}$ is the correlation matrix pertaining to $\hat{\Sigma}$, and the notation $\underset{\text{approx.}}{\sim}$ indicates the approximate distribution for large $n$. The family of hypotheses from \eqref{testing} can then equivalently be expressed as  
\begin{equation}
\label{test}
{H_{0j}: \mu_{j} = 0 } \text{~~versus~~} {H_{1j}: \mu_{j} \neq 0 }, \thickspace j = 1, \cdots, p.
\end{equation}

\subsection*{Approximation of the false discovery proportion}
Throughout this manuscript, we consider the multiple test problem which is given by the $p$ pairs of null and alternative hypotheses specified in \eqref{test}. Let $p_{0} = \#\{j: \mu_{j} = 0\}$ denote the number of true null hypotheses and $p_{1} = \#\{j: \mu_{j} \neq 0\}$ the number of false null hypotheses, such that $p = p_{0}+ p_{1}$. For the calibration of a multiple test with respect to type I error control, we proceed similarly as in Storey's method (see \cite{storey2002}). 
 Namely, for a (data-dependent) threshold ${t}$, we will reject the null hypotheses which correspond to those p-values that are not exceeding $t$. This approach has been broadly used in practice (e.\ g., see, \cite{pfa}, \cite{fan_poet}, \cite{efron2007, efron2010}, \cite{storey2002}).
  The aim of the proposed method is to estimate the realized FDP for any fixed $t$ in the multiple testing setting given by \eqref{test}, based on the Z-statistics \eqref{zstat} under an arbitrary structure of $\Sigma$.

To this end, we consider empirical processes given by
\begin{align*}
V(t) &= \#\{\text{true null~} P_{j}: P_{j} \leq t\},\\
S(t) &= \#\{\text{false null~} P_{j}: P_{j} \leq t\},\\
R(t) &= \#\{P_{j}: P_{j} \leq t\},
\end{align*}
where ${t}$ ranges in $[0, 1]$. 
\begin{table}[!ht]
\centering
	\caption{\bf Decision pattern of the multiple test which thresholds the marginal $p$-values at a given value $t \in [0, 1]$.}\label{tab:hyp_table}
	\begin{tabular}{c c c c}
		\hline 
		Number & Number accepted  & Number rejected & Overall  \\   
		\hline
		True nulls   & $U(t)$    & $V(t)$ & $p_{0}$  \\
		False nulls & $T(t)$  & $S(t)$    & $p_{1}$ \\
		All nulls & ${p}-R(t)$  & $R(t)$    & ${p}$ \\
		\hline
	\end{tabular}
\end{table}
 For a given value of $t$, the null hypothesis $H_{0, j}$ is rejected if and only if its corresponding $p$-value $p_{j}$ does not exceed $t$. This decision rule leads to the decision pattern which is displayed in Table \ref{tab:hyp_table}. The random variables $V(t)$, $S(t)$, and $R(t)$ are the number of false discoveries (i.\ e., false rejections), the number of true discoveries and the total number of discoveries, respectively. Clearly, $R(t) = V(t) + S(t)$. The latter random variables depend on the test statistics $Z_{1}, Z_{2}, \ldots, Z_{p}$, because every $p$-value $P_j$ is a transformation of the corresponding $Z$-statistic $Z_j$, $1 \leq j \leq p$, as we will describe below. Furthermore, $V(t)$ and $S(t)$ are both unobservable, whereas $R(t)$ is observable. We recall here the definition of the FDP, namely, $\text{FDP}(t) = V(t) / \max\{R(t), 1\}$.
 
\subsection*{Principal Factor Approximation}
 The next step of the analysis is to model and to utilise the dependency structure of the test statistics in an approximation of FDP$(t)$ for a given $t$. The proposed technique relies on an approximation of a normally distributed random vector with a factor model involving weakly dependent, normally distributed random errors. To this end, we first employ a spectral decomposition of the correlation matrix $\hat{\Sigma}^*$ (cf. \cite{pfa}).
 Namely, $\hat{\Sigma}^*$ is represented in terms of its eigenvalue-eigenvector pairs $(\lambda_j, \gamma_j)_{1 \leq j \leq p}$, where $\lambda_{1} \geq \lambda_{2} \geq \cdots \geq \lambda_{p} \geq 0$. The representation can be written as
 \begin{equation} \label{pfa}
 \hat{\Sigma}^* = \lambda_{1}\gamma_{1}\gamma_{1}^{\top} + \lambda_{2}\gamma_{2}\gamma_{2}^{\top} + \cdots + \lambda_{p}\gamma_{p}\gamma_{p}^{\top}.
 \end{equation}
For a fixed integer $k \geq 1$,  we let $A_k = \sum_{j = k + 1}^{p} \lambda_{j}\gamma_{j}\gamma_{j}^{\top}$, and we note that
 \begin{equation} 
 	\norm{A_k}_{F}^{2} = \lambda_{k + 1}^{2} + \cdots \lambda_{p}^{2}, 
 \end{equation}
 where $\norm{.}_F$ is the Frobenius norm. We further let $L_k = (\sqrt{\lambda_{1}}\gamma_{1}, \sqrt{\lambda_{2}}\gamma_{2}, \cdots,  \sqrt{\lambda_{k}}\gamma_{k})$, which presents a $p \times k$ matrix. Thus, $\hat{\Sigma}^*$ can be written as,  
 \begin{equation}
 \hat{\Sigma}^* = L_k L_k^{\top} + A_k.
 \end{equation}
Respectively, $Z_{1}, \ldots, Z_{p}$ can be expressed as
\begin{equation}
\label{pfa_eq}
Z_{j} = \mu_{j} + b_{j}^{\top} W + K_{j} = \mu_{j} + \eta_j + K_{j}, \quad j = 1, \cdots, p,
\end{equation}
where $b_{j} = (b_{j1}, \ldots, b_{jk})^{\top}$ and  $(b_{1j}, \ldots, b_{pj})^{\top} = \sqrt{\lambda_{j}}\gamma_{j}$. The vector $W = (W_{1}, \ldots, W_{k})^\top \sim N_k(0, I_k)$ is called the vector of factors,  and these factors are stochastically independent of each other. The random vector $(K_{1}, \cdots ,K_{p})^\top \sim N_p(0, A_k)$ is called the vector of random errors, and  it is assumed that factors and random errors are stochastically independent. We can think of \eqref{pfa_eq} as a model for the data-generating process for $Z_1, \ldots, Z_p$. In this interpretation, $\mu_j = 0$ corresponds to the true null hypotheses, and $\mu_j \neq 0$ corresponds to the false null hypotheses.

It is essential to choose the number ${k}$ of factors carefully. One the one hand, it is important to choose $k$ large enough to capture most of the dependencies among $Z_1, \ldots, Z_p$. On the other hand, a small $k$ stabilizes the computations, both from a numerical and from a statistical point of view. In \cite{pfa}, one way to determine a suitable value of $k$ has been discussed. Concretely, the authors proposed to choose the smallest $k$ such that
 \begin{equation}
 \label{epsilon}
\frac{\sqrt{\lambda_{k+1}^{2} + \cdots + \lambda_{p}^{2}}}{\lambda_{1} + \cdots+ \lambda_{p}} < \epsilon,
\end{equation}
where $\epsilon$ is some small number, for example, $0.01$.
It has been pointed out in \cite{fan_poet}  that an overestimation of $k$ does not invalidate the approximation of the FDP, as long as the unobserved factors can still be estimated with a reasonable accuracy.

Based on the aforementioned derivations, we consider the "principal factor" FDP estimator from Proposition 2 in  \cite{pfa}, 
which is given by
 \begin{equation}
 \label{pfa_est}
 \widehat{\text{FDP}}(t) = \min \left\{ \sum_{j =  1 }^{p} \left[\Phi(a_{j}(z_{t/2} + \widehat{\eta}_{j}) + \Phi(a_{j}(z_{t/2} - \widehat{\eta}_{j}))\right], R(t)\right\} / R(t)
 \end{equation}
 whenever $R(t) \neq  0$, and $\widehat{\text{FDP}}(t) = 0$ in the case of $R(t) = 0$. In \eqref{pfa_est}, $a_j = (1 - \sum_{h = 1}^{k} b_{jh}^2)^{-1/2}$ and $R(t) = \{j: 2\Phi(-|Z_{j}|) \leq t \}$ is the (total) number of  rejections for a given $t$, where
 $\Phi$ and $z_{t/2} = \Phi^{-1}(t/2)$ are the cumulative distribution function and the lower $t/2$-quantile of the standard normal distribution on $\mathbb{R}$, respectively. The (unadjusted) two-sided (random) $p$-value corresponding to $Z_j$ is given by $P_j = 2\Phi(-|Z_{j}|) = 2 (1 - \Phi(|Z_j|))$, and this $p$-value is thresholded at $t$ for every $j \in \{1, \ldots, p\}$ when computing $R(t)$. Furthermore, $\hat{\eta}_{j} = \sum_{h = 1}^{k} b_{jh}\widehat{W}_{j}$ is an estimator for $\eta_{j} = b_{j}^\top W$.

The estimator in \eqref{pfa_est} relies on the intuition that large $|\mu_{j}|$'s tend to generate large $|z_{j}|$'s, meaning that false null hypotheses tend to produce large Z-statistics (in absolute value). Furthermore, the estimator in \eqref{pfa_est} relies on a sparsity assumption, namely, that the number $p_0$ of true null hypotheses is close to $p$. This assumption justifies the summation over all $j$ from one to $p$ in \eqref{pfa_est}. Different FDP estimators have been compared in \cite{fdp_vs_fdr}, and under sparsity in the aforementioned sense the author has proposed to use the estimator from \eqref{pfa_est}.
There are several reasons why the assumption of sparsity is plausible in our study. Firstly, due to high sensitivity during sample preparation and acquisition, there is evidence of a small signal-to-noise ratio. Secondly, a reasonable assumption is that solely a tiny fraction of molecular masses are distinctive for a cancer association. In fact, we have applied the proposed method to real MALDI data, where there have been characterised five biomarkers (i.\ e., biologically meaningful covariates) out of a couple of thousands of measured covariates. 

In order to evaluate \eqref{pfa_est} in practice, it remains to specify the estimator
$\widehat{W} = (\widehat{W}_{1}, \ldots, \widehat{W}_{k})^\top$ of the common factors. 
In \cite{pfa}, it has been proposed to construct $\widehat{W}$ by means of 
$L_2$-regression or by means of $L_1$-regression, respectively. For the former, the authors proposed to include only the 90\% smallest $|z_j|$'s in the regression fit. 
Specifically, we denote by 
$w = (w_{1}, \ldots , w_{k})^{\top}$ the realized values of $\{W_{h}\}_{h=1}^{k}$, and by $\hat{w}$ the estimator for $w$. Then, the estimator based on $L_2$-regression is given by
\begin{equation}
\label{ols}
\hat{w} = \min_{W} \sum_{ j = 1}^{ \lfloor 0.9 p \rfloor } (Z_{j} - b_{j}^\top W)^{2},
\end{equation}
where we assume that the $Z_j$'s in \eqref{ols} are ordered from small to large according to their absolute values.
This estimator has been used in our simulation study. The estimator based on $L_1$-regression is given by
\begin{equation}
\label{l1}
\hat{w} = \min_{W} \sum_{ j = 1}^{p} |Z_{j} - b_{j}^{\top}W|.
\end{equation}
We adopted $L_1$-regression rather than $L_2$-regression, because it is more robust to outliers. 

Finally, the dependency-adjusted (random) $p$-values corresponding to the $Z_j$'s are given by  
\begin{equation}
\label{adjmethod}
\tilde{P}_{j} = 2\Phi(-|a_{j}(Z_{j} - b_{j}^{\top}\widehat{W})|).
\end{equation} 
The null hypothesis $H_{0j}$ from \eqref{test} gets rejected based on the observed data, iff  $\tilde{p}_{j} \leq {t}$, $1 \leq j \leq p$. In this, the data-dependent rejection threshold is chosen as the largest value $t = t_\alpha \in [0, 1]$ such that $\widehat{\text{FDP}}(t_\alpha)$ is not exceeding a pre-defined level $\alpha$. In practice, a (grid) search algorithm can be employed to find the value $t_\alpha$ for a given level $\alpha$.

\subsection*{Schematic description of the entire data analysis workflow}
Algorithm \ref{algo:statmodel1} provides a step-by-step description of the proposed  data analysis workflow.
\begin{footnotesize}
\begin{algorithm}
	\begin{algorithmic}[1]
		\STATE 	Fit the marginal logistic regression model with the logit link function for each $j \in \{1, \ldots, p\}$ separately on the basis of  $(X_{j}^{(i)}: 1 \leq i \leq n, 1 \leq j \leq p)$ and $(Y^{(i)}: 1 \leq i \leq n)$.\linebreak
		\onehalfspacing  (1.1) Find the maximum-likelihood estimates for $\hat{\beta}_j$ and $\hat{\alpha}_j$.\\ 
		\onehalfspacing (1.2) Calculate the standardized score contributions $\hat{\Psi}_{ij}$ based on $\hat{\beta}_{j}$ for $i \in \{1, \ldots, n\}$ and stack them on top of each other to build a  vector $\hat{\Psi}_{i}$.\\
		\onehalfspacing(1.3) Calculate the estimated covariance matrix $\hat{\Sigma}$, given in \eqref{mmmvariance}, based on  $(\hat{\Psi}_{i}: 1 \leq i \leq n)$, and obtain the correlation matrix pertaining to $\hat{\Sigma}$. \\	
		\onehalfspacing(1.4) Calculate the Z-statistics given in \eqref{z_values}.
		\STATE Based on the Z-statistics, evaluate $R(t)$ for a given threshold $t$. 
		\STATE Apply singular value decomposition to the correlation matrix pertaining to $\hat{\Sigma}$, and determine an appropriate number of factors ${k}$. Then, extract the corresponding factor loading coefficients $\{b_{jh}: j = 1, \ldots, p;  h = 1, \ldots, k\}$.
		\STATE Obtain the estimates $\widehat{W}_{1}, \cdots, \widehat{W}_{k}$ of the common factors by means of regression; cf.\ \eqref{ols} and \eqref{l1}, respectively. Plug these factor estimates into \eqref{pfa_est}, and obtain the estimate $\widehat{\text{FDP}}(t)$, for a given $t$.
		\STATE  Obtain adjusted $p$-values according to \eqref{adjmethod}. 
		\STATE Threshold the adjusted $p$-values at $t_\alpha$ for FDP control at a given level $\alpha$.
	\end{algorithmic}
\caption{The Logit-PFA method}
	\label{algo:statmodel1}
\end{algorithm}
\end{footnotesize}

\subsection*{MALDI imaging data}
We applied the proposed multiple testing approach to a MALDI IMS data frame introduced in \cite{krieg2}. In \cite{krieg2}, five biomarkers have been characterized (see \textit{Supplementary Table 1} in \cite{krieg2}).
  Broadly speaking, biomarkers are biologically meaningful molecules indicative of a distinct biological state or condition (cf. \cite{biomarker}). Statistically speaking, biomarkers are well-identified predictors that can be used to accurately predict relevant clinical outcomes, and also, they are an apt starting point for an evaluation of any statistical model. 
  The aforementioned data frame has been re-analyzed by several researchers; cf.\ \cite{Boskamp}, \cite{johannes}, and \cite{Behrmann}. Therefore, we refer to the aforementioned references for an extensive description of the data frame. Here, we only give a brief overview of sample acquisition, data preparation, measurement and data processing. 
  
  FFPE lung tumour tissues samples, for this study, were provided by the bank of the National Center for Tumour Diseases (NCT, Heidelberg, Germany). Cylindrical tissue cores of non-small cell lung cancer were taken from 304 patients, where 168 patients were associated with primary lung adenocarcinoma (ADC), and 136 patients were associated with primary squamous cell carcinoma (SqCC). Cylindrical tissue cores of all tissue samples were collected in eight TMA blocks in total. Lung cancer is the leading reason for cancer-related deaths worldwide, with around 1.59 million reported deaths in 2012 (for more concrete numbers, see, e.\ g., \cite{krieg2} or \cite{nscl}). Two primary lung cancer categories are determined, namely small cell lung cancer and non-small cell lung cancer (NSCLC), whence the latter constituted around 85\% of all cases. The two most fatal histological NSCLC entities are ADC and SqCC, compromising of approx. 50\% and approx. 40\% of all lung cancers, respectively. Differentiation of these two subtypes is critical for the choice of chemotherapy regimens and further test strategies.
  
  Tissue sections were cut from all TMA blocks and treated in accordance with a previously published protocol for tryptic peptide imaging; cf.\ \cite{protocol}. MALDI data were obtained through an Autoflex speed MALDI-TOF instrument (Bruker Daltonik) in positive ion reflector mode. Spectra were measured in the mass range 500–5000 m/z at 150 µm spatial resolution using 1600 laser shots. Tumour status and typing for all cores were confirmed by standard histopathological examination; cf.\ \cite{Boskamp}.
   Afterwards, the raw spectral data was loaded into SCiLS Lab (version 2016b, Bruker Daltonik), the standard baseline correction was performed (convolution method of 20), and total-ion-count (TIC) normalization was employed.  
  The normalising step is crucial in order to reduce the laboratory variation resulting from day-to-day instrument fluctuations or biological artefacts coming from sample preparation. 
  Finally, spectral smoothing was performed to intervals of 1 Da (dalton) width (cf.\ \cite{senko}),
   and the spectra were pruned to the mass range of 500-3545 m/z values (outside this interval m/z values were not considered), which resulting in 3046 m/z channels (columns).  
   
   In summary, we worked on a MALDI data set where all data-processing steps are based on standard protocols. It is out of the scope of this paper to compare different data-processing steps, like normalisation, smoothing, etc.

\section*{Results}
\subsection*{Simulation Studies}
In this section, we illustrate the performance of the proposed approach based on simulated data under different data-generating processes. Specifically, we consider the sample size $n = 400$, the number of false nulls hypotheses $p_{1} = 10$, and the total number of hypotheses $p \in \{500, 1000\}$. For each combination of these parameters, $1{,}000$ simulation runs have been performed. For a given value of $p_1$, we assume without loss of generality that $\beta_{j} \neq 0$ for $j \in \{1, \ldots, p_1\}$, while the $p_0$ true nulls with $\beta_{j} = 0$ correspond to the coordinates $j \in \{p_1+1, \ldots, p\}$. We employed the least-squares estimator, defined in \eqref{ols}, for the estimation of $\{\widehat{W}_{h} : h = 1, \ldots, k\}$. 
   For each observational unit $i \in \{1, \ldots, n\}$ and each coordinate (or: covariate) $j \in \{1, \ldots, p\}$, we consider the model 
   \[\mathbb{P}_\beta(Y_{i} = 1 | X_{j}^{(i)} ) = \frac{\exp(X_{j}^{(i)}\beta_{j})}{1 + \exp(X_{j}^{(i)}\beta_{j})}
   \]
   for the response variable $Y_i$ given the covariate $X_{j}^{(i)}$, meaning that all intercepts $\alpha_j$ have been set to zero for $j \in \{1, \ldots, p\}$. In our simulations, we have moreover set $\beta_j = 1$ for all $j \in \{1, \ldots, p_1\}$.
   The considered data-generating distributions for the vector $X = (X_1, \ldots, X_p)^\top$ are provided in Model \ref{model-simu}.
   
   \begin{model}\label{model-simu} $ $\\
   \onehalfspacing\underline{Scenario 1:} $X_{1}, \ldots, X_{p}$ are stochastically independent and identically $N(0, 1)$-distributed random variables. \\
   \onehalfspacing\underline{Scenario 2:} 
   $X_{1}, \ldots, X_{p}$ are jointly normally distributed on $\mathbb{R}^p$. The parameters of their $p$-variate joint normal distribution have been chosen  such that each $X_j$ is marginally $N(0, 1)$-distributed, $j = 1, \ldots, p$.
   Furthermore, the correlation coefficient $\text{Corr}(X_{j_1}, X_{j_2})$ equals $\rho$ for all $1 \leq j_1 < j_2 \leq p_1$ as well as for all $p_1+1 \leq j_1 < j_2 \leq p$ (Gaussian equi-correlation model). The subvector $(X_j: 1 \leq j \leq p_1)$ is stochastically independent of the subvector $(X_j: p_1+1 \leq j \leq p)$, to avoid spurious effects of covariates $X_j$ with $p_1 +1 \leq j \leq p$ on the response variable which arise from confounding of covariates $X_j$ with $1 \leq j \leq p_1$.\\
   \onehalfspacing\underline{Scenario 3:} 
As Scenario 2, but now $(X_j: 1 \leq j \leq p_1)$ are stochastically independent and identically $N(0, 1)$-distributed random variables.
    \end{model}
    
For the simulation of correlated independent variables, we have used the  function \verb=rmvnorm= from the \verb=R= package \verb=mvtnorm=. In Tables \ref{table:sim_1_1} - \ref{table:sim_3_2}, we report summaries (over the $1{,}000$ simulation runs) of $\widehat{\text{FDP}}(t)$, $R(t)$, and $S(t)$ for fixed values of $t$, and we report the median value of $t_\alpha$ for the common choice of $\alpha = 0.05$.

Tables \ref{table:sim_1_1} - \ref{table:sim_1_2} summarize our simulation results under Scenario 1. Here, due to joint independence of the test statistics, $t_{0.05}$ is rather small, because the "effective number of tests" (in the sense of Section 3.4 in \cite{handbook-chapter} and the references therein) equals $p$ under joint independence of the test statistics, meaning that a rather strong multiplicity correction is required. On the other hand, the standard error of  $\widehat{\text{FDP}}(t)$ is rather small under Scenario 1, too, because the FDP concentrates well around its expectation (the FDR) under joint independence of all $p$ test statistics. The results given in Tables \ref{table:sim_2_1} - \ref{table:sim_2_2} refer to Scenario 2 of Model \ref{model-simu}, and Tables \ref{table:sim_3_1} - \ref{table:sim_3_2} refer to Scenario 3. Under Scenarios 2 and 3, the effective number of tests is smaller than $p$ whenever $\rho > 0$, and it decreases with increasing $\rho$. Thus, $t_{0.05}$ increases with $\rho$, too. Under our Scenario 2, the considered multiple test always rejected all ten false null hypotheses. For this reason, we do not report summaries of $S(t)$ in Tables \ref{table:sim_2_1} and \ref{table:sim_2_2}. The reason for the high power of the multiple test under Scenario 2 is, that the correlation among $(X_j: 1 \leq j \leq p_1)$ amplifies the signal strength for each $j \in \{1, \ldots, p_1\}$. Under Scenario 3, where the relevant regressors are stochastically independent, the power of the multiple test is smaller than under Scenario 2, such that on average only approximately seven of the ten false null hypotheses can be rejected by the multiple test considered in Tables \ref{table:sim_3_1} and \ref{table:sim_3_2}.

\begin{table}[!ht]
	\begin{adjustwidth}{-2.25in}{0in}
		\caption{\bf Simulation results under Scenario 1 (I)}
		\begin{tabular}{@{\extracolsep{0.5pt}}lccccccc}
			\hline \\
	\shortstack{Median of\\ $\widehat{\text{FDP}}(t)$} & \shortstack{ Standard Error \\ of $\widehat{\text{FDP}}(t)$}   &\shortstack{Average of\\ $R(t)$} & \shortstack{ Standard Error \\ of $R(t)$}  & \shortstack{Average of \\   $S(t)$} & \shortstack{ Standard Error \\ of $S(t)$} & \shortstack{Median of $t_{0.05}$}   \\
			\hline\\
		0.004144 &  0.000918 & 6.930 & 1.247 & 6.892  & 1.236 & 1.24e-03\\ 
			\hline \\
		0.009116 & 0.002175  & 6.965  &  1.270 & 6.898 &  1.228 &  6.6e-04\\  
			\hline
		\end{tabular}
			\begin{flushleft}The total number of hypotheses equals $p = 500$ in the first row and $p=1{,}000$ in the second row;\\  
			the number of factors equals $k = 10$; the rejection threshold equals $t = 10^{-4}$, except for the last column.
		\end{flushleft}
		\label{table:sim_1_1}
	\end{adjustwidth}
\end{table}
\begin{table}[!ht]
	\begin{adjustwidth}{-2.25in}{0in}
		\caption{\bf Simulation results under Scenario 1 (II)}
		\begin{tabular}{@{\extracolsep{0.5pt}}lcccccc}
			\hline \\
			\shortstack{Median of\\ $\widehat{\text{FDP}}(t)$} & \shortstack{ Standard Error \\ of $\widehat{\text{FDP}}(t)$}   &\shortstack{Average of\\ $R(t)$} & \shortstack{ Standard Error \\ of $R(t)$}  & \shortstack{Average of \\   $S(t)$} & \shortstack{ Standard Error \\ of $S(t)$}   \\
			\hline\\
			0.158180 &  0.022796 & 11.774 & 1.664 & 9.519  & 0.632 \\ 
			\hline \\
			0.277878 & 0.045845  & 14.062  &  2.194 & 9.517 &  0.650  \\  
			\hline
		\end{tabular}
			\begin{flushleft}The total number of hypotheses equals $p = 500$ in the first row and $p=1{,}000$ in the second row;\\ 
			the number of factors equals $k = 10$; the rejection threshold equals $t = 0.005$, except for the last column.
	\end{flushleft}
		\label{table:sim_1_2}
	\end{adjustwidth}
\end{table}


\begin{table}[!ht]
	\begin{adjustwidth}{-2.25in}{0in}
	\caption{\bf Simulation results under Scenario 2 (I)}
	\begin{tabular}{@{\extracolsep{0.5pt}}lcccccc}
		\hline \\
		\shortstack{$\rho$} & \shortstack{Median of\\ $\widehat{\text{FDP}}(t)$} & \shortstack{ Standard Error \\ of $\widehat{\text{FDP}}(t)$}   &\shortstack{Average of\\ $R(t)$} & \shortstack{ Standard Error \\ of $R(t)$} & \shortstack{Median of $t_{0.05}$}   \\
		\hline\\
		0.2 & 0.001395 & 0.009750 & 10.041  & 0.203 &  2.41e-03 \\   
		\hline \\
		0.5 &0.000131 & 0.024349 & 10.031   & 0.173 &  7.41e-03\\    
		\hline \\
		0.8 &0.000099 & 0.018323 & 10.031   & 0.173  & 3.49e-02\\    
		\hline	
	\end{tabular}
	\begin{flushleft}The total number of hypotheses equals $p = 500$;  the number of factors equals $k = 1$;\\ 
	the rejection threshold equals $t = 10^{-4}$, except for the last column.
	\end{flushleft}
	\label{table:sim_2_1}
	\end{adjustwidth}
\end{table}

\begin{table}[!ht]
	\begin{adjustwidth}{-2.25in}{0in}
	\caption{\bf Simulation results under Scenario 2 (II)}
	\begin{tabular}{@{\extracolsep{0.5pt}}lccccc}
		\hline \\
	\shortstack{$\rho$} &\shortstack{Median of\\ $\widehat{\text{FDP}}(t)$} & \shortstack{ Standard Error \\ of $\widehat{\text{FDP}}(t)$}   &\shortstack{Average of\\ $R(t)$} & \shortstack{ Standard Error \\ of $R(t)$} &\shortstack{Median of $t_{0.05}$} \\
		\hline\\
	0.2	&0.002842 & 0.014190 & 10.086  & 0.298  & 1.28e-03  \\   
		\hline \\
	0.5	&0.000167 & 0.028907 & 10.076   & 0.2689  & 4.69e-03\\    
		\hline \\
	0.8	&0.000099 & 0.024967 & 10.073   & 0.271  & 2.74e-02 \\    
		\hline	
	\end{tabular}
	\begin{flushleft}The total number of hypotheses equals $p = 1000$;  
	the number of factors equals $k = 1$;\\ 
	the rejection threshold equals $t = 10^{-4}$, except for the last column.
	\end{flushleft}
	\label{table:sim_2_2}
		\end{adjustwidth}
\end{table}

\begin{table}[!ht]
	\begin{adjustwidth}{-2.25in}{0in}
		\caption{\bf Simulation results under Scenario 3 (I)}
		\begin{tabular}{@{\extracolsep{0.5pt}}lccccccc}
			\hline \\
	\shortstack{$\rho$} &\shortstack{Median of\\ $\widehat{\text{FDP}}(t)$} & \shortstack{ Standard Error \\ of $\widehat{\text{FDP}}(t)$}   &\shortstack{Average of\\ $R(t)$} & \shortstack{ Standard Error \\ of $R(t)$}  & \shortstack{Average of \\   $S(t)$} & \shortstack{ Standard Error \\ of $S(t)$} & \shortstack{Median of $t_{0.05}$} \\
			\hline\\
		0.2 &0.002292 &   0.012646 & 6.930 & 1.250 & 6.902  & 1.240 & 2.41e-03 \\ 
			\hline \\
		0.5 &0.000221 & 0.026264  &  6.924  &  1.251 & 6.898 &  1.240  & 7.41e-03 \\  
			\hline \\
			 0.8 &0.000142 & 0.021452  &  6.917  &  1.248 & 6.895 &  1.237 &  3.49e-02 \\  
			\hline
		\end{tabular}
			\begin{flushleft}The total number of hypotheses equals $p = 500$;  the number of factors equals $k = 1$;\\ the rejection threshold equals $t = 10^{-4}$, except for the last column.
		\end{flushleft}
		\label{table:sim_3_1}
	\end{adjustwidth}
\end{table}

\begin{table}[!ht]
	\begin{adjustwidth}{-2.25in}{0in}
		\caption{\bf Simulation results under Scenario 3 (II)}
		\begin{tabular}{@{\extracolsep{0.5pt}}lcccccccc}
			\hline \\
	\shortstack{$\rho$} &\shortstack{Median of\\ $\widehat{\text{FDP}}(t)$} & \shortstack{ Standard Error \\ of $\widehat{\text{FDP}}(t)$}   &\shortstack{Average of\\ $R(t)$} & \shortstack{ Standard Error \\ of $R(t)$}  & \shortstack{Average of \\   $S(t)$} & \shortstack{ Standard Error \\ of $S(t)$} & \shortstack{Median of $t_{0.05}$}   \\
			\hline\\
	0.2	&0.004194 &  0.021576 &6.937 & 1.298 & 6.872  & 1.272 & 1.28e-03  \\ 
			\hline \\
	0.5	&0.000250 & 0.039067  &  6.936  &  1.30 & 6.870 &  1.273  & 4.69e-03 \\  
			\hline \\
	0.8		&0.000143 & 0.024121  &  6.940  &  1.30 & 6.877 &  1.273  & 2.74e-02 \\  
			\hline
		\end{tabular}
			\begin{flushleft}The total number of hypotheses equals $p = 1000$; the number of factors equals $k = 1$;\\ 
			the rejection threshold equals $t = 10^{-4}$, except for the last column.
		\end{flushleft}
		\label{table:sim_3_2}
	\end{adjustwidth}
\end{table}
 
\subsection*{Analysis of real MALDI imaging data}
\begin{figure}[!ht]
\includegraphics[width=\textwidth]{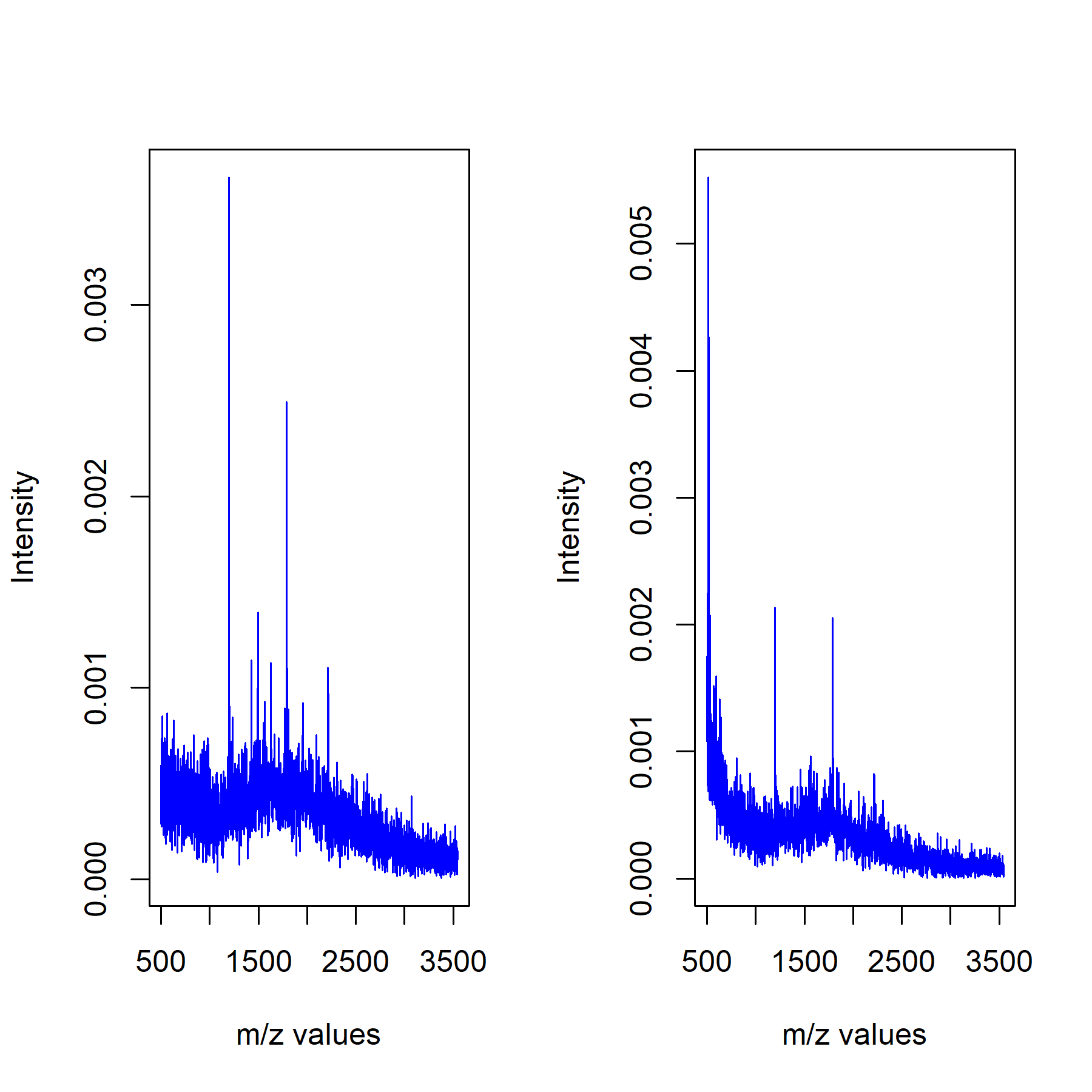}
\caption{\textbf{Two exemplary MALDI spectra.} Two unique and specific spots within a tissue. Each of these spots represent a mass spectrum.}
\label{fig:examplespectra}
\end{figure}

Figure \ref{fig:examplespectra} displays two mass spectra from both cancer subtypes, where the m/z values are illustrared on the horizontal axes, while  the vertical axes refer to the relative abundances (intensities values) of ionizable molecules. These two graphs represent unique and specific spots within a patient's tissue, and correspond to two mass spectra. We, therefore, model each pixel marginally to identify which m/z values (based on 1 DA) are distinctive for a particular cancer subtype. We refer to \cite{Behrmann} (see their Figure 1) for a more detailed illustration of the pipeline from a tissue to a single spectrum.   

\begin{figure}[!ht]
\includegraphics[width=\textwidth]{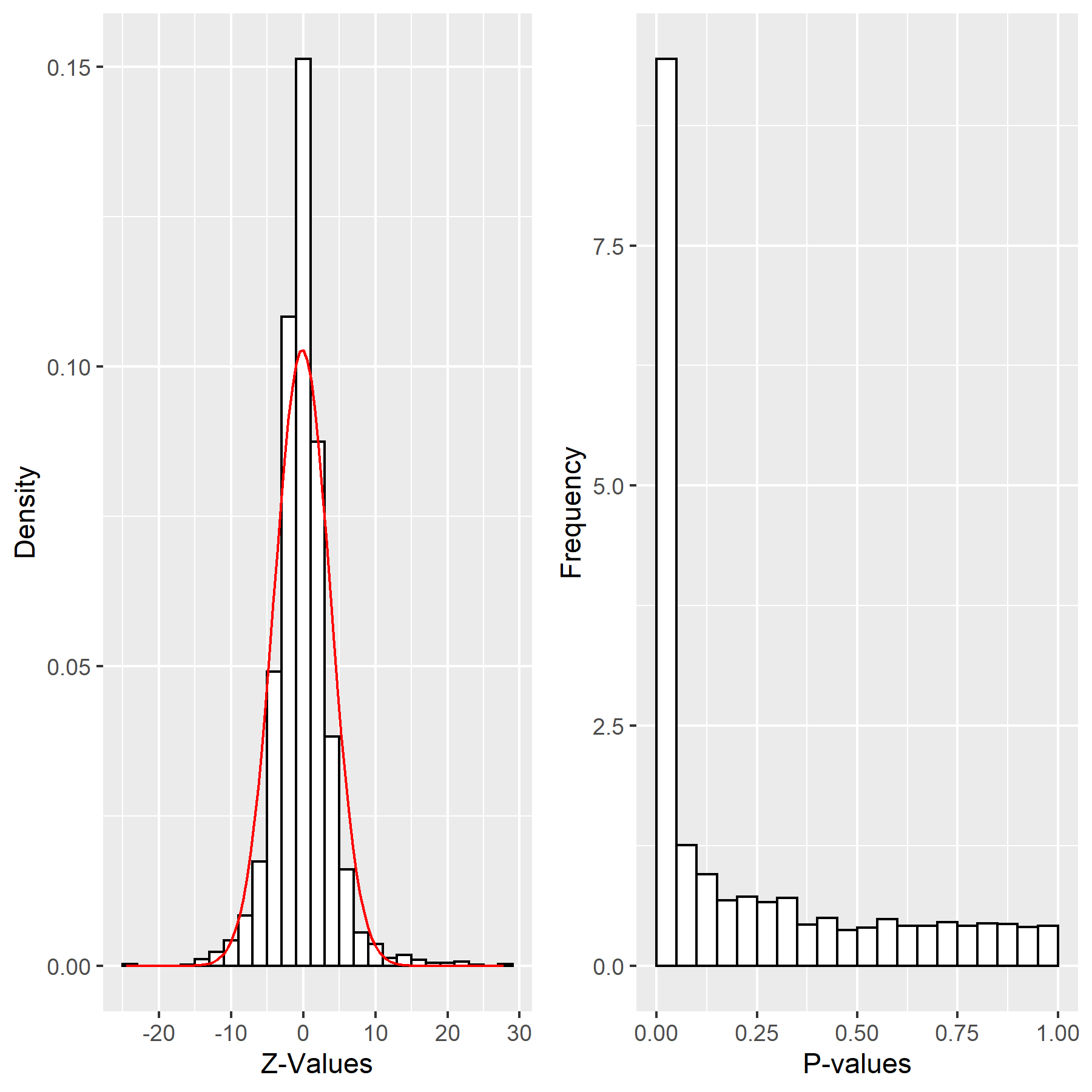}
\caption{\textbf{The empirical distribution and fitted normal density curve of the Z-values for the MALDI data.} Due to dependencies among 
 			the $Z$-values, they are not following the theoretical $N(0, 1)$ distribution. Instead, a closer look at the empirical distribution reveals that it can best be approximated by $N(-0.115, 3.88^2)$. Consequently, the non-adjusted p-values have a lot of mass around zero.}
 \label{fig:z_p_values}
 \end{figure}

As discussed in \cite{efron2007} and \cite{efron2010}, the density of the empirical distribution of all $Z$-values does in general not coincide with the density of the standard normal distribution on $\mathbb{R}$, even if almost all $p$ null hypotheses are true. The reason for this phenomenon, which can also be observed on our data (see Figure \ref{fig:z_p_values}), is the presence of dependencies among the $Z$-statistics. In particular, these dependencies lead to an inflation of the variance of the null distribution of the $Z$-statistics. However, we nevertheless have that the $Z$-statistics of the previously identified biomarkers lie in tails of the distribution. Namely, their $Z$-statistics are large in absolute value, and might be declared as statistically significant. Note that we consider an absolute value for the $Z$-values, since we wish to find distinctive m/z values for either cancer subtype. 

\begin{figure}[!ht]
	\includegraphics[width=\textwidth]{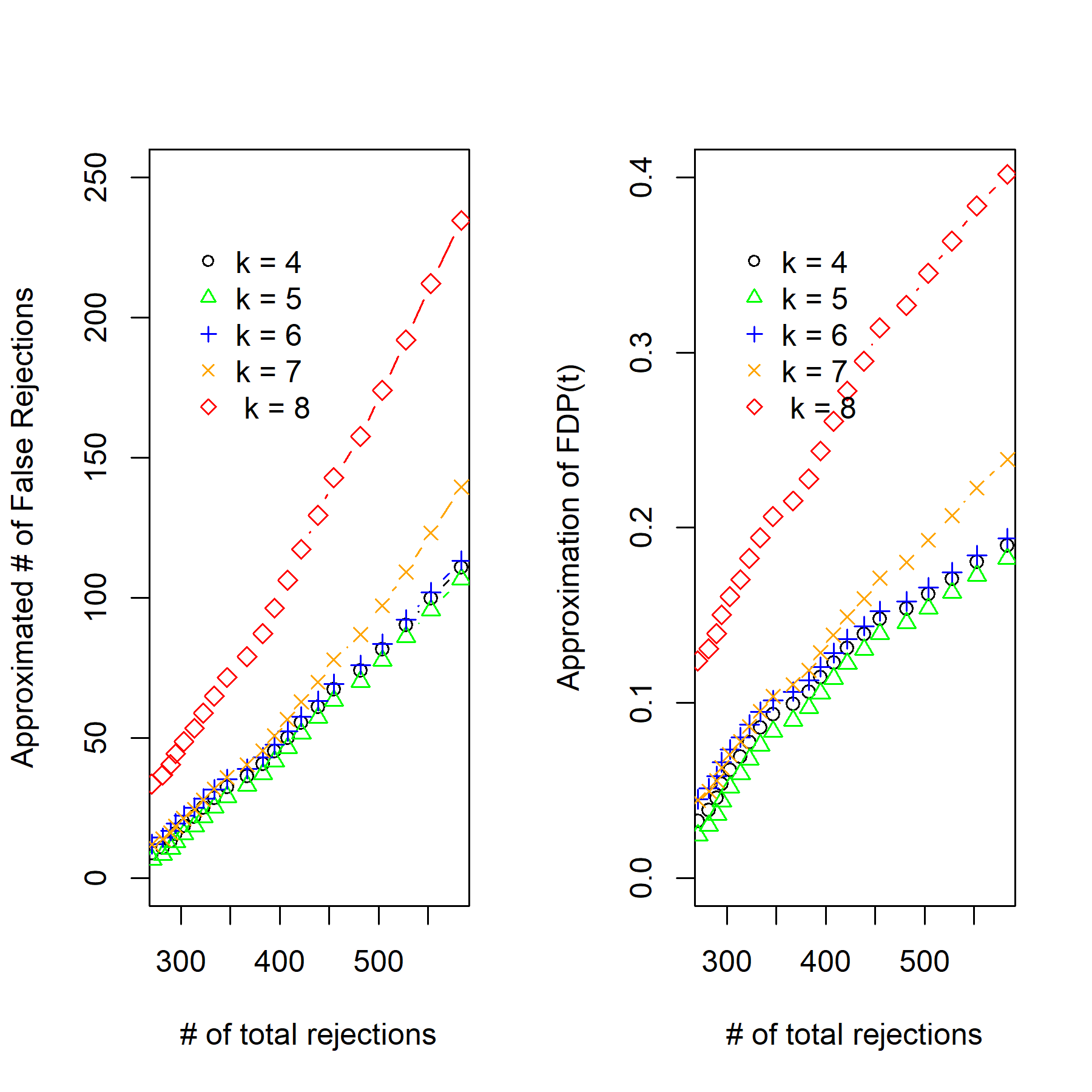}
 \caption{\textbf{The approximated number of false discoveries as well as the approximated FDP as functions of the total number of rejections.} Each curve corresponds to a different choice of the number $k$ of common factors, where  $k \in \{4, 5, 6, 7, 8\}$ has been considered.}
\label{fig:comparison}
\end{figure}

  The next step of the data analysis has been to determine an appropriate number $k$ of common factors. To this end, we performed the proposed data analysis workflow described in Algorithm \ref{algo:statmodel1} over a range of different candidate values for $k$ and compared the results. As documented in Figure \ref{fig:comparison}, the estimated number of false discoveries as well as the estimated FDP remain rather stable for $k \leq 7$. Based on this, we chose $k = 6$ for our actual data analysis.  
  
\begin{figure}[!ht]
\includegraphics[width=\textwidth]{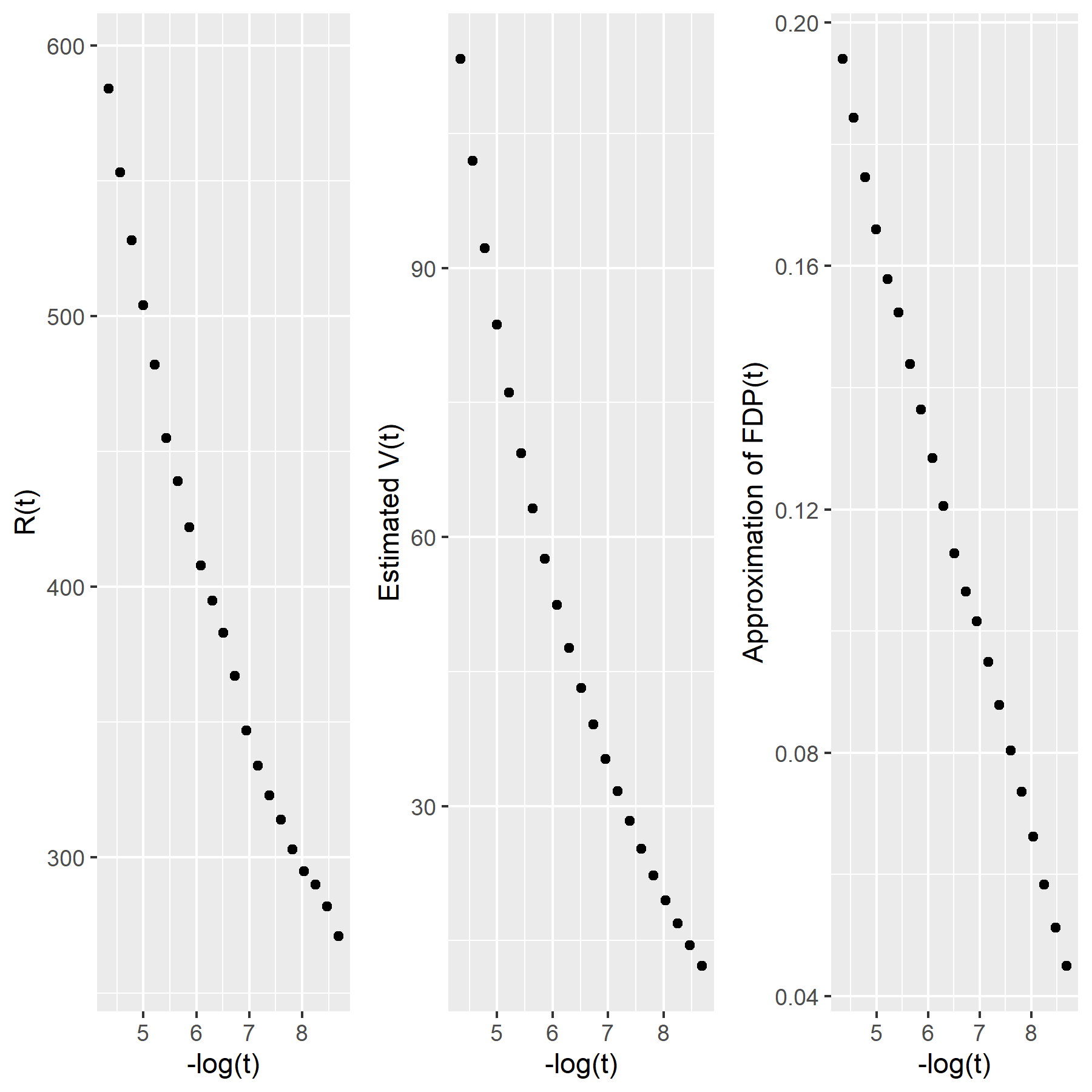}
\caption{\textbf{Main results: Total number of rejections, estimated number of false rejections, and estimated FDP, as functions of the threshold $t$.} For the sake of presentation, the values of the $t$-axis are provided on a negative logarithmic scale.}
	\label{fig:results}
\end{figure}

The main results of our real data analysis are illustrated in Figure \ref{fig:results}. It is evident that $R(t)$, $\widehat{V}(t)$ and $\widehat{\text{FDP}}(t)$ are increasing in the rejection threshold $t$. For $t \in [10^{-9}, 10^{-5}]$, the estimated FDP lies between 4 \% and 20\%. This indicates that most of the smallest $p$-values correspond to false nulls (leading to true discoveries). 

However, the absolute magnitude of the observed  $Z$-values is grossly large. Therefore, we picked the data-driven threshold for actual decision making rather small. Table \ref{table:rangeoftvalues} lists the total number of rejections as well as the estimated FDP for several plausible choices of $t$.

\begin{table}[!ht]
\caption{\bf Number of rejections and estimated FDP for several plausible rejection thresholds.}
	\centering
	\begin{tabular}{rrr}
		\hline
		 Threshold $t$ &$R(t)$ &$\widehat{\text{FDP}}(t)$  \\ 
		\hline
		 1.01e-05 & 504 & 0.1659 \\
		 1.37e-06 & 422  & 0.1364 \\
		 3.06e-07 & 383 & 0.1127\\
		 1.13e-07 & 347 & 0.1016\\ 
		 9.24e-09 & 295 & 0.066\\ 
		 2.06e-09 & 271  & 0.0445\\ 
		\hline
	\end{tabular}
 \label{table:rangeoftvalues}
\end{table}

The Logit-PFA method indicates, as highly significant, m/z values that are closely related to the five biomarkers identified in \cite{krieg2}, for all  considered thresholds $t$. These findings were confirmed by the dependency-adjustment method and also by the original $Z$-statistics with a fixed threshold value. Table \ref{table:z10} lists the 15 top-ranked (i.\ e., most significant) null hypotheses (m/z values) for both cancer subtypes and thereby illustrates the overall significance of the five previously identified biomarkers which are indicated by stars in Table \ref{table:z10}. For a comparison to the findings published previously in \cite{krieg2}, \cite{Boskamp} and \cite{johannes}, we attribute the values m/z = 1411 and m/z = 1412 to the peak of a peptide of the CK5 protein (m/z = 1410.7) and its second isotopic peak. In addition, the values m/z = 1878 and m/z = 1907 appearing in  Table \ref{table:z10} are likely to be attributable to peptides of the proteins CK15 (monoisotopic m/z = 1877.9) and HSP27 (monoisotopic m/z = 1905.9). The value m/z = 1407, indicating a negative direction, is likely associated to a peptide of the CK7 protein (monoisotopic m/z = 1406.7), which indicates to be efficient for an identification for ADC in the lung. The last peak at m/z = 1822 can be attributed to a peptide CK15 protein (monoisotopic m/z = 1821.9) distinctive for SqCC.

\begin{table}[!ht]
	\caption{\bf Top 15 ranked m/z values based on their original Z-values for both cancer subtypes.} 
	\begin{subtable}[t]{.5\textwidth}
		\caption{The most sign. m/z-values for ADC}
		\raggedright
		\begin{tabular}{@{\extracolsep{0pt}}lcc}
			\hline\\
			m/z values & \multicolumn{1}{c}{Z-values} & \multicolumn{1}{c}{P-values}  \\
			\hline\\
			1407* & -24.53  & $<10^{-6}$ \\
			1408 & -23.59  & $<10^{-6}$    \\
			654  & -16.50     & $<10^{-6}$     \\
			1235 & -14.32     & $<10^{-6}$     \\
			1813 & -13.79     & $<10^{-6}$    \\
			1706 & -13.63   & $<10^{-6}$    \\
			2247 & -13.49      & $<10^{-6}$    \\
			1477 & -13.23    & $<10^{-6}$    \\
			1814 & -13.25    & $<10^{-6}$    \\
			2855 & -13.13    & $<10^{-6}$    \\
			1517 & -12.82    & $<10^{-6}$    \\
			1812 & -12.64    & $<10^{-6}$    \\
			2246 & -12.57    & $<10^{-6}$    \\
			1294 & -12.49    & $<10^{-6}$    \\
			1278 & -12.48    & $<10^{-6}$    \\ 
			\hline \\
		\end{tabular}
	\end{subtable}%
	\begin{subtable}[t]{.5\textwidth}
		\raggedleft
		\caption{The most sign. m/z-values for SqCC}
		\begin{tabular}{@{\extracolsep{0pt}}lcc}
			\hline\\
			m/z values & \multicolumn{1}{c}{Z-values} & \multicolumn{1}{c}{P-values}  \\
			\hline\\
			1412 &  27.74  & $<10^{-6}$  \\
			1411* &  27.59  &$<10^{-6}$    \\
			811  &  24.63     & $<10^{-6}$     \\
			1878* &  22.39     & $<10^{-6}$     \\
			1879 & 22.14     & $<10^{-6}$    \\
			866 & 22.01   & $<10^{-6}$    \\
			1822* & 21.81      & $<10^{-6}$    \\
			879 & 20.64    & $<10^{-6}$    \\
			1823 & 20.63    & $<10^{-6}$    \\
			1413 & 19.12    & $<10^{-6}$    \\
			1880 & 18.99    & $<10^{-6}$    \\
			1907 & 18.83    & $<10^{-6}$   \\
			1906* & 18.80    & $<10^{-6}$    \\
			1908 & 16.90    & $<10^{-6}$    \\
			1426 & 16.37    & $<10^{-6}$   \\ 
			\hline \\
		\end{tabular}
	\end{subtable}
	\begin{flushleft}
	Those m/z value that are presumably related to the five previously identified biomakers are indicated by the symbol * in both subtables.
	\end{flushleft}
	\label{table:z10}
\end{table}

Finally, Table \ref{table:proportions} displays the distribution of significant findings (for the plausible threshold values listed in Table \ref{table:rangeoftvalues}) over different mass ranges.
The distributions given in Table \ref{table:proportions} are in good agreement with the results in Figure 4 of \cite{Behrmann}. Namely, the authors of \cite{Behrmann} pointed out that most signals appeared in the mass range of 1000-2000 m/z. This is expected due to the fact that most peptides are measured within this array, and our analysis indicates a similar behaviour of the false null hypotheses within particular ranges of m/z values. Moreover, in the regime $[2800, 3545]$ of very high m/z values  there were 6 to 9 significant m/z values across the choices of $t$ given in the first column of Table \ref{table:proportions}. This is also in line with the results in Figure 4 of  \cite{Behrmann}.

\begin{table}[!ht]
\begin{adjustwidth}{-2.25in}{0in}
	\caption{\bf Distribution of significant findings across mass ranges.} 
	\begin{tabular}{@{\extracolsep{1pt}}lccc}
		\hline\\
		Threshold  & Mass Range $\le$ 1000 m/z & Mass Range $\in$ (1000, 2000] m/z & Mass Range $>$ 2000  m/z \\ 
		\hline \\
		 1.01e-05 & 26.11  &  56.48  &  17.41  \\
		1.37e-06 & 25.38  & 58.79  & 15.84  \\
		3.06e-07 &  24.94 & 60.10 & 14.96 \\
		1.13e-07 & 24.94  &  60.45 & 14.61  \\ 
		9.24e-09 & 24.20  &  62.97 & 12.83  \\ 
		2.06e-09 & 24.54  &  63.19 & 12.27  \\ 
		\hline
	\end{tabular}
	\begin{flushleft}
	Significance is based on the proposed dependency-adjusted procedure.
	\end{flushleft}
	\label{table:proportions}
\end{adjustwidth}	
\end{table}

\section*{Discussion}
From the statistical perspective, we have proposed an inferential framework for two-sample comparisons in high-dimensional settings when the test statistics have an arbitrary correlation structure. The major assumptions underlying the proposed methodology are (i) asymptotic normality of the vector of test statistics and (ii) that the dependency structure among the test statistics can be described accurately by a factor model. To account for the high multiplicity of the considered applications, our criterion for type I error control is to bound the false discovery proportion. This also accounts for strong dependencies among test statistics, because in that case the FDP is typically not well concentrated around its mean (the FDR), and hence many authors have considered FDP control as the more appropriate criterion than FDR control under strong dependencies; see, e.\ g., \cite{sinica2014} and the references therein.  

From the application perspective, we have applied the proposed method to a MALDI imaging data frame with a large number of covariates (m/z values). The results derived with the proposed method are consistent with already reported insights about this data frame. However, to the best of our knowledge we have for the first time contributed a statistically grounded significance evaluation to the empirical findings. Reliable statistical modelling of MALDI data is a challenging task; cf., e.\ g., \cite{jonathan}. Our approach based on MMM does not rely on heavy assumptions. Essentially, it is assumed that the (binary) phenotype of interest is associated with certain m/z-values, and that this association can be described by a (marginal) logistic regression model for each m/z-value separately. These assumptions are well established in the statistical theory of modelling binary data; see, e.\ g., \cite{agresti}.

There are several possible directions for future research: First, it may be interesting to consider other supervised statistical learning models (for instance, neural networks with more than one layer) instead of the logistic regression model proposed in this work. Second, it is of interest to quantify the uncertainty about the realized FDP for different threshold values, with the goal of providing a confidence region for this realized FDP, in addition to a mere point estimate. Third, it is of interest to analyze the statistical properties of MALDI data in a more detailed manner, which may allow for a joint modelling (after potential dimension reduction) instead of the MMM-based approach presented here. Finally, it will be worthwhile to consider categorical response variables with more than two categories.

\section*{Acknowledgments}
The authors acknowledge the aid of Jonathan von Schroeder for a highly detailed and kind introduction to the MALDI data. Also, for many meaningful and fruitful discussions related to the applicability of multiple testing to the MALDI context. We thank J. Leuschner for his constructive criticism of the manuscript.
 The authors appreciatively acknowledge the financial support from the German Research Foundation within the framework of RTG 2224, entitled "$\pi^{3}$: Parameter Identification—Analysis, Algorithms, Applications".

\nolinenumbers

%
%
%

\end{document}